\documentclass[twocolumn,amsmath,floatfix,prl,aps]{revtex4-1}

\usepackage{color}
\usepackage{lipsum,amsmath}
\usepackage{pifont}   
\usepackage{graphicx} 
\usepackage{dcolumn}  
\usepackage{bm}       
\usepackage{amsfonts} 
\usepackage{amssymb}  
\usepackage{multirow} 
\usepackage{tikz}
\usepackage{siunitx}
\usepackage{placeins}
\usepackage{braket}
\usepackage{nicefrac}
\usepackage{bm}
\usepackage{physics}
\usepackage{grffile}

\usepackage{array,multirow,graphicx}
\usepackage{ulem}
\usepackage[latin1]{inputenc}
\usepackage{caption}
\usepackage{scrextend}
\usepackage{hyperref}
\usepackage{float}
\usepackage{subcaption}
\usepackage{gensymb}

\usepackage[english]{babel}
\usepackage[autostyle, english = american]{csquotes}
\MakeOuterQuote{"}

    \renewcommand{\v}[1]{\bm{\mathrm{#1}}}
    \newcommand{\m}[1]{\bm{\mathsf{#1}}}

\begin{document}

\title{Where does the spin angular momentum go in laser induced demagnetisation?}
\author{J.~K. Dewhurst$^1$}
\author{S. Shallcross$^2$}
\author{P. Elliott$^2$}
\author{S. Eisebitt$^{2,3}$}
\author{C. v. Korff Schmising$^2$}
\author{S. Sharma$^2$}
\email{sharma@mbi-berlin.de}
\affiliation{1 Max-Planck-Institut fur Mikrostrukturphysik Weinberg 2, D-06120 Halle, Germany}
\affiliation{2 Max-Born-Institute for Non-linear Optics and Short Pulse Spectroscopy, Max-Born Strasse 2A, 12489 Berlin, Germany}
\affiliation{3 Institut f\"ur Optik und Atomare Physik, Technische Universit\"at Berlin, 10623 Berlin, Germany.}

\date{\today}

\begin{abstract}
The dynamics of ultrafast demagnetisation in 3$d$ magnets is complicated by the presence of both spin ${\v S}$ and orbital ${\v L}$ angular momentum, with the microscopic mechanism by which the magnetic moment is redistributed to the lattice, and at what time scales, yet to be resolved. Employing state-of-the-art time dependent density function theory we disentangle the dynamics of these two momenta. Utilising ultra short (5~fs) pulses
that separate spin-orbit (SO) and direct optical excitation time scales, we demonstrate a two-step microscopic mechanism: (i) an initial loss of ${\v L}$ due to laser excitation, followed post pulse by (ii) an increase of ${\v L}$ as ${\v S}$ transfers to ${\v L}$ during subsequent ($> 15$~fs) SO induced spin-flip demagnetisation. We also show that to see an unambiguous transfer of ${\v S}$ to ${\v L}$ a short pulse is required.
\end{abstract}

\maketitle


A universal feature of itinerant magnets is that they can be demagnetized on femtosecond time scales by application of ultrashort laser pulses\cite{beaurepaire_ultrafast_1996,Boeglin2010,bovensiepen_femtomagnetism:_2009}. However, more than two decades since the demagnetisation of matter by laser light was first demonstrated a fundamental question remains unanswered: where does the spin angular momentum go?

Angular momentum conservation implies that the loss of the magnetic moment must ultimately result in an increase in angular momentum of the lattice, and unambiguous evidence of this can be seen in the lattice waves that occur on sub-picosecond time scales\cite{dornes_ultrafast_2019}.
However, while the ultimate fate of the spin moment is not in doubt the microscopic mechanism of its transfer to the lattice at ultrafast time scales remains unclear.
In particular, as the spin moment ${\v S}$ does not couple directly to the lattice, its transfer to the lattice must be mediated by the spin-orbit coupling term, predominately through the orbital momentum ${\v L}$ of the electron system. On this basis one would expect to see an increase in ${\v L}$ in the early time spin dynamics, resulting from the transfer of angular momentum from spin  to orbital degree of freedom via spin-orbit coupling. Confounding this expectation, all experiments to date find no increase in orbital angular momentum\cite{Boeglin2010,Bergeard2014,Hennecke2019a,Stamm2010}.

In pump probe experiments spin and orbital angular momentum are determined from magnetic circular dichroism (MCD) and X-ray absorption spectra (XAS), using sum rules involving excitation from core or semi-core states\cite{thole,carra,altarelli,kunes,chen}. The limited time-resolution of these experiments cannot exclude faster angular momentum transfer processes, in particular the anticipated flow of angular momentum from the spin moment via the orbital moment to the lattice. Indeed, theoretically within a tight-binding model it was noted that orbital angular momentum is rapidly reduced already at very short 
time scales ($L_z$ is quenched and transferred to the lattice $<0.1$~fs\cite{Tows2015}, where $z$ is the spin quantization axis), possibly explaining the absence of any observed increase in ${L_z}$ in experiment.
However, as the dynamics of ${\v L}$ will be driven by optically induced currents one would expect the essential time scale of the ${\v L}$ dynamics to be controlled by the pulse full width half maximum (FWHM). The dynamics of the spin angular momentum, on the other hand, is primarily governed by the spin-orbit (SO) coupling times. A separation between these two times can thus be affected by pulse control, potentially conclusively demonstrating the early time flow of the angular momentum from one sub-system to another. 
Employing time-dependent density functional theory (TDDFT) we show that the time scales of ${\v L}$ and ${\v S}$ can be controlled and, crucially, separated by using ultra short pulses that separate the SO and direct optical excitation times. On this basis we subsequently demonstrate that early time loss in ${\v L}$ is accompanied by a later time increase, as ${\v S}$ transfers to ${\v L}$, resolving the long standing mystery of how the spin moment of magnetic order ${\v S}$ transfers to the lattice.


{\it Methodology:} we will employ time dependent density functional theory which has proven to be a powerful tool for investigating early time spin dynamics. It has been used to predict the phenomena of optically induced spin transfer leading to control over magnetic structure at ultrashort time scales\cite{dewhurst_laser-induced_2018}, subsequently observed in several experiments\cite{siegrist_light-wave_2019,steil-heusler,hofherr-feni,clemens-copt,chen_competing_2019}, and is able to very accurately describe moment loss in bulk 3d element magnets\cite{krieger_laser-induced_2015,SSBM17} and thin magnetic overlayers\cite{PhysRevApplied.10.044065}.

According to the Runge-Gross theorem\cite{RG1984}, which extends the Hohenberg-Kohn theorem of the ground state into the time domain, for common initial states there will be a one-to-one correspondence between the time-dependent external potentials and densities\cite{Carsten_book}.
Based on this theorem, a system of non-interacting particles can be chosen such that the density of this non-interacting system is equal to that of the interacting system for all times, with the wave function of this non-interacting system represented by a Slater determinant of single-particle orbitals. In a fully non-collinear spin-dependent version of these theorems\cite{KDES15} the time-dependent Kohn-Sham (KS) orbitals are Pauli spinors governed by the Schr\"odinger equation:

\begin{eqnarray}
i\frac{\partial \psi_j({\bf r},t)}{\partial t}&=&
\Bigg[
\frac{1}{2}\left(-i{\nabla} +\frac{1}{c}{\bf A}_{\rm ext}(t)\right)^2 +v_{s}({\bf r},t)+ \nonumber \\
&&\frac{1}{2c} {\m \sigma}\cdot{\bf B}_{s}({\bf r},t) + \nonumber \\
&&\frac{1}{4c^2} {\m \sigma}\cdot ({\nabla}v_{s}({\bf r},t) \times -i{\nabla})\Bigg]
\psi_j({\bf r},t)
\label{KS}
\end{eqnarray}
where ${\bf A}_{\rm ext}(t)$ is a vector potential representing the applied laser field, and ${\m \sigma}$ the vector of Pauli matrices $(\sigma_x,\sigma_y,\sigma_z)$. The KS effective potential $v_{s}({\bf r},t) = v_{\rm ext}({\bf r},t)+v_{\rm H}({\bf r},t)+v_{\rm xc}({\bf r},t)$ is decomposed into the external potential $v_{\rm ext}$, the classical electrostatic Hartree potential $v_{\rm H}$ and the exchange-correlation (XC) potential $v_{\rm xc}$, for which we have used the adiabatic local density approximation. Similarly the KS magnetic field is written as ${\bf B}_{s}({\bf r},t)={\bf B}_{\rm ext}(t)+{\bf B}_{\rm xc}({\bf r},t)$ where ${\bf B}_{\rm ext}(t)$ is the external magnetic field and ${\bf B}_{\rm xc}({\bf r},t)$ is the exchange-correlation (XC) magnetic field. The final term of Eq.~\eqref{KS} is the spin-orbit coupling term. 
From the Kohn-Sham states the response can then be calculated\cite{PRL-felix,PRL-dewhurst,clemens-copt}. To obtain an accurate response function, needed to extract ${\v L}$ and ${\v S}$ using sum rules, requires very high convergence, for which a  $20\times20\times20$ grid of $\v k$-vectors and excited states up to 120~eV above the Fermi energy were used. Local field effects were included in calculation of the response function. 

In a density function treatment how does angular momentum flow to the lattice?
The change in the electronic charge density due to the pump laser pulse generates forces on the atomic nuclei. These forces, which will include the angular analogue of well known linear Hellmann-Feynman forces, will then generate motion of the lattice. In the present work the back reaction on the electron system from subsequent nuclear displacement is neglected; it is not expected to be significant at the ultrashort time scales considered here.

\begin{figure}[h]
{\includegraphics[width=\columnwidth]{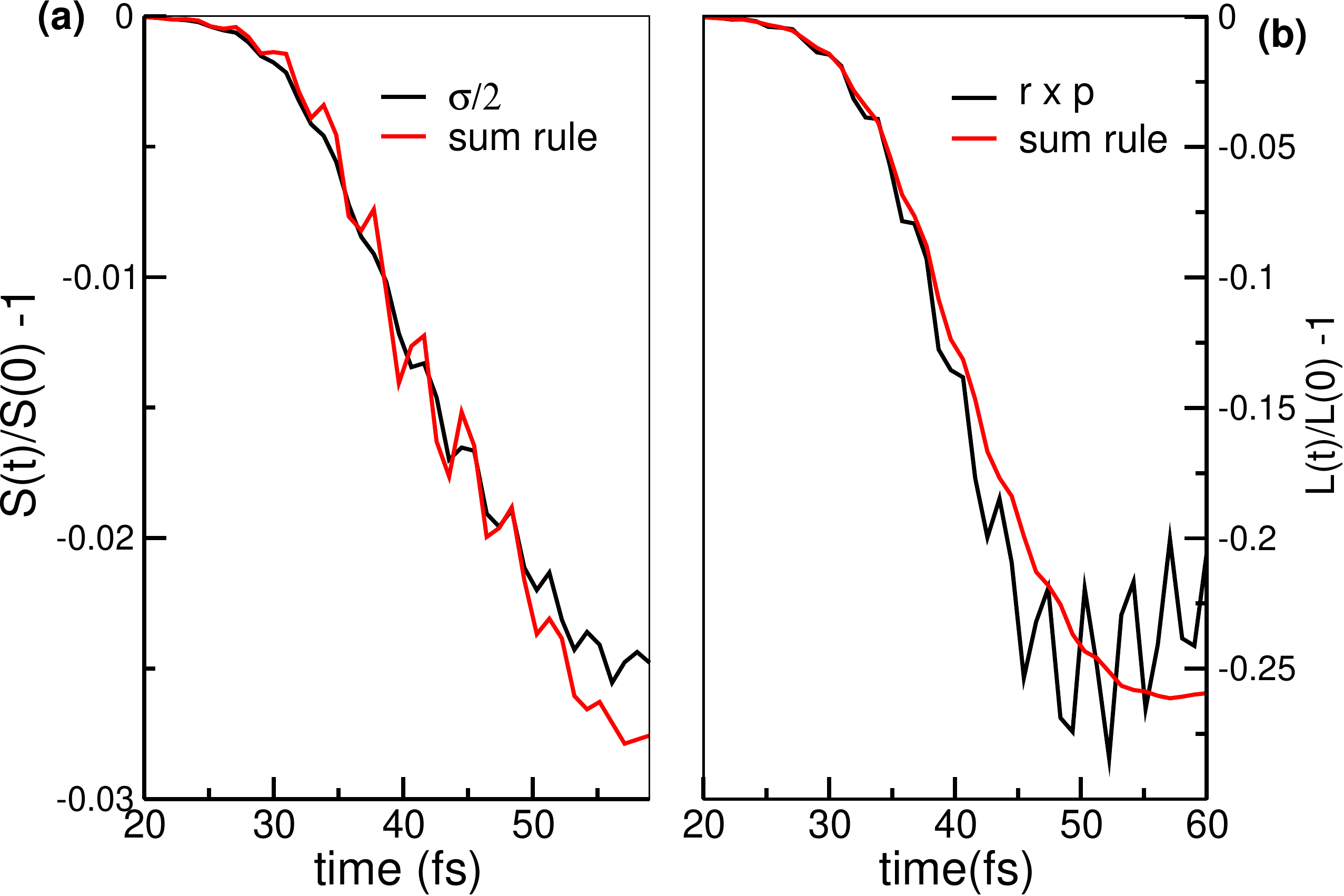}}
\caption{Angular momentum obtained from the optical response and L-edge sum rules and directly from the $\v S$ and $\v L$  operators. Shown are (a) the normalized spin angular moment $S_z(t)/S_z(0)-1$ and (b) the normalized orbital angular momentum  $L_z(t)/L_z(0) -1$, for Co as a function of time in femtoseconds (fs). A pump pulse with central frequency of 1.55eV, fluence 2~mJ/cm$^2$, and full width half maximum 20~fs was used.}\label{f:sumrules}
\end{figure}
%

\begin{figure}[h]
\includegraphics[width=\columnwidth]{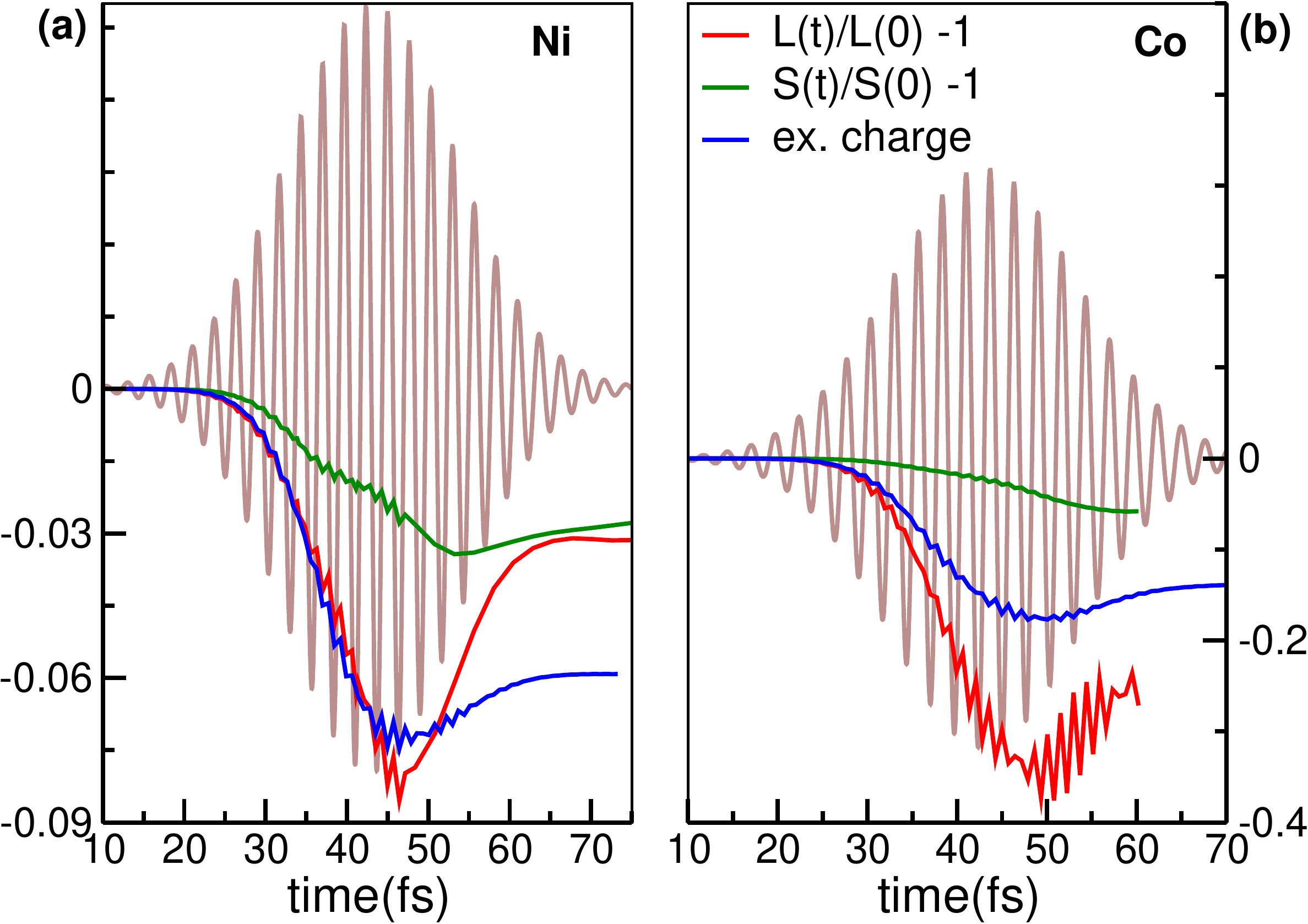}
\caption{Spin dynamics of $L_z$ and $S_z$ for a laser pulse of full width half maximum (FWHM) 25~fs, central frequency 1.55eV and incident fluence 3.55mJ/cm$^2$. Shown are the normalized orbital angular momentum, $L_z(t)/L_z(0)-1$, and spin angular moment, $S_z(t)/S_z(0)-1$ for (a) Ni and (b) Co. The excited charge and the $A$-field of the pump pulse are also shown.}
\label{fig:longp}
\end{figure}

{\it Results:} rigorously speaking, in a periodic solid the position operator $\v r$ is not well defined, and so the orbital angular momentum operator $\v L = \v r \cross \v p$ also cannot be defined. However, if the angular momentum arises from orbitals localized within the muffin-tin (a sphere around the point nuclei), which in our calculations are chosen to be as large as possible i.e. touching, then this approach can still employed: one sets ${\v r} =0$ as the centre of the muffin-tin and performs the integral of the $\v L$ expectation value only within this sphere. For the transient case the basic assumption is then that the current loops induced by the laser are predominantly contained within the muffin-tins (for an in depth discussion of these current loops see Ref.~\onlinecite{elliott_microscopic_2020}). A very different method for calculating the spin $S_z$ and orbital angular momentum $L_z$ is to extract them from optical response of core/semi-core states via well known sum rules\cite{thole,carra,altarelli,kunes,chen}. In this approach the assumption that current loops reside only within the muffin-tins is relaxed, as the states in the entire unit-cell are used to determine the response function. This is the experimental method of choice as the optical response function can be efficiently obtained via XMCD and XAS in a pump probe experiment. However, as discussed by Resta\cite{PhysRevResearch.2.023139} for periodic systems this approach is not formally guaranteed to yield the exact physical spin and orbital angular momenta.

To clarify this unsettling situation we will establish numerically that these two distinct methodologies provide consistent information. To deploy the sum rule method requires additional work in that not just the static but also the transient response function must be calculated accurately. This can be done and the results thus calculated have been shown to be in excellent agreement with the experimental data, as has been demonstrated in Refs.~\onlinecite{PRL-felix,PRL-dewhurst,clemens-copt}.

In Fig.~\ref{f:sumrules} we compare the spin and angular momentum calculated by (i) applying the sum rules\cite{thole,carra,altarelli,kunes,chen} to the calculated transient XMCD and XAS spectra at the L-edge and, (ii) from the expectation of the $\v S$ and $\v L$ operators calculated from the Kohn-Sham wavefunctions and integrated within the muffin-tin sphere about each atom. As can be seen, angular momentum derived from the response function agrees very well with direct evaluation from the corresponding operators. It is especially notable that ${S_z}$ extracted from the spin sum rule $S_z=n_h(3p-2q)/r$ agrees very well with the magnetization vector field integrated in a sphere around the nuclei, as this sum rule involves differences in $q$ and $p$ making it more sensitive to the optical spectra than the $L_z$ sum rule $L_z=n_h 4q/3r$ ($q$ and $p$ are the integral of the spectrum and the integral of the $L_3$ edge respectively, $r$ the XAS integrals, and $n_h$ the number of holes in the $d$-band\cite{thole,carra,altarelli,kunes,chen}). 
Furthermore, we note that the change in the orbital and spin angular momentum obtained using the (ii) method is entirely along the $z$-axis.
The consistent information provided by these different methods of obtaining the ${\v S}$ and ${\v L}$ angular momentum resolves any doubts concerning their veracity: for the experimentally relevant fluences (2-5~mJ/cm$^2$ incident fluence) employed here the muffin-tin integrated angular momentum expectation values may be used in full confidence that they will correspond with experimentally measured angular momenta. The origin of this agreement likely resides both in the low fluence and the localized nature of $p$ and $d$ orbitals, which are almost completely circumscribed by the muffin-tin sphere. The oscillations in ${L_z}$ are exactly twice the period of the pump pulse and are the result of excited charge oscillating in and out of the muffin tin, as shown in Ref.~\onlinecite{elliott_ultrafast_2016}.

{\it Long pulse}: with this agreement between the two methods for obtaining the angular momentum we now address the question of why an increase in $L_z$ is not seen in experiments. We first simulate the electron dynamics induced by an optical pulse with a duration of 25 fs (FWHM), approximately a factor of two  shorter than used in the most recent experiments exploring the separate response of spin and orbital angular momenta during ultrafast demagnetization \cite{Hennecke2019a}. 
In Fig.~\ref{fig:longp}, the time resolved spin and orbital angular momentum are shown for Ni and Co. In both cases we observe ultrafast demagnetization, with a $3\%$ in Ni, and $1\%$ in Co, loss of the spin moment. The fact that for a pulse of fixed fluence Ni demagnetizes more than Co agrees very well with the experimental observations\cite{Koopmans2010}.

For both Co and Ni, the orbital angular momentum initially decreases, which can be understood as the ground-state angular momentum, caused by spin-orbit splitting of the $d$-shell levels, being lost due to the optical excitation of electrons. After the maximum of the pulse is reached the orbital moment can then be seen to increase. This change, however, cannot be assigned solely to an angular momentum flow from the spin to the orbital degree-of-freedom, as during the interval of increasing $L_z$ the electron system is de-exciting, as may be seen from the number of excited electrons\cite{PhysRevB.89.064304} (defined as the charge above $E_f$ in the ground-state minus the charge above $E_f$ at any given time, where $E_f$ is the ground state Fermi energy). Hence for this long pulse we cannot unambiguously discern a transfer of the magnetic moment from the spin to the orbital degree-of-freedom. Furthermore, during this 50-100~fs time window electron-electron and electron-phonon scattering will become active, complicating the dynamics and acting to further reduce ${\v S}$ and ${\v L}$.
In short, all that may be concluded from these results is that optical excitations generate a very short time reduction in ${\v L}$, and so subsequent ${\v S}$ to ${\v L}$ transfer may not result in a measurable increase in ${L_z}$.
The dynamically changing $\v L$ implies currents, consequent re-distribution of charge, and the generation of forces on the nuclei that would set the lattice in motion, accomplishing the transfer of angular momentum from the magnetisation to the lattice.

\begin{figure}[t]
\includegraphics[width=\columnwidth, clip]{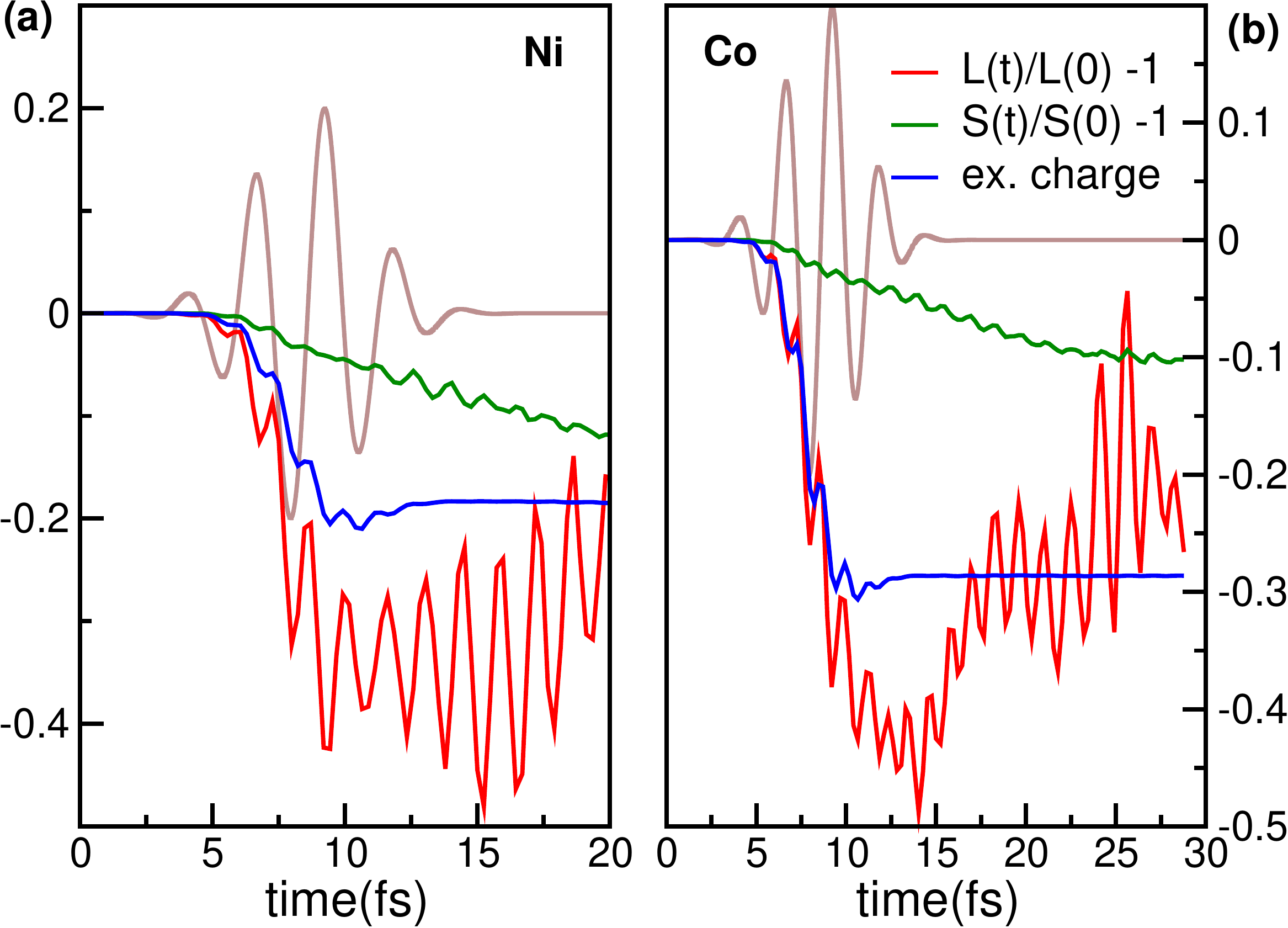}
\caption{Dynamics of spin and angular momentum for an ultrashort pulse of full width half maximum 5~fs. Shown are the normalized orbital angular momentum, $L_z(t)/L_z(0)-1$, and spin angular moment, $S_z(t)/S_z(0)-1$ for (a) Ni and (b) Co. The $A$-field of the pump pulse and excited charge are also presented. The pump pulse has an incident fluence of 5mJ/cm$^2$ and central frequency 1.55eV. Note that in contrast to the longer pulse of FWHM 25~fs shown in Fig.~\ref{fig:longp}, significant change in $L_z$ and $S_z$ can be seen post pulse during which no change in the amount of excited charge is seen. This results from the continuous loss of $S_z$ into the lattice via the orbital angular angular momentum $L_z$ of the electron system.}\label{fig:shortp}
\end{figure}
%

\begin{figure}[t]
\includegraphics[width=\columnwidth, clip]{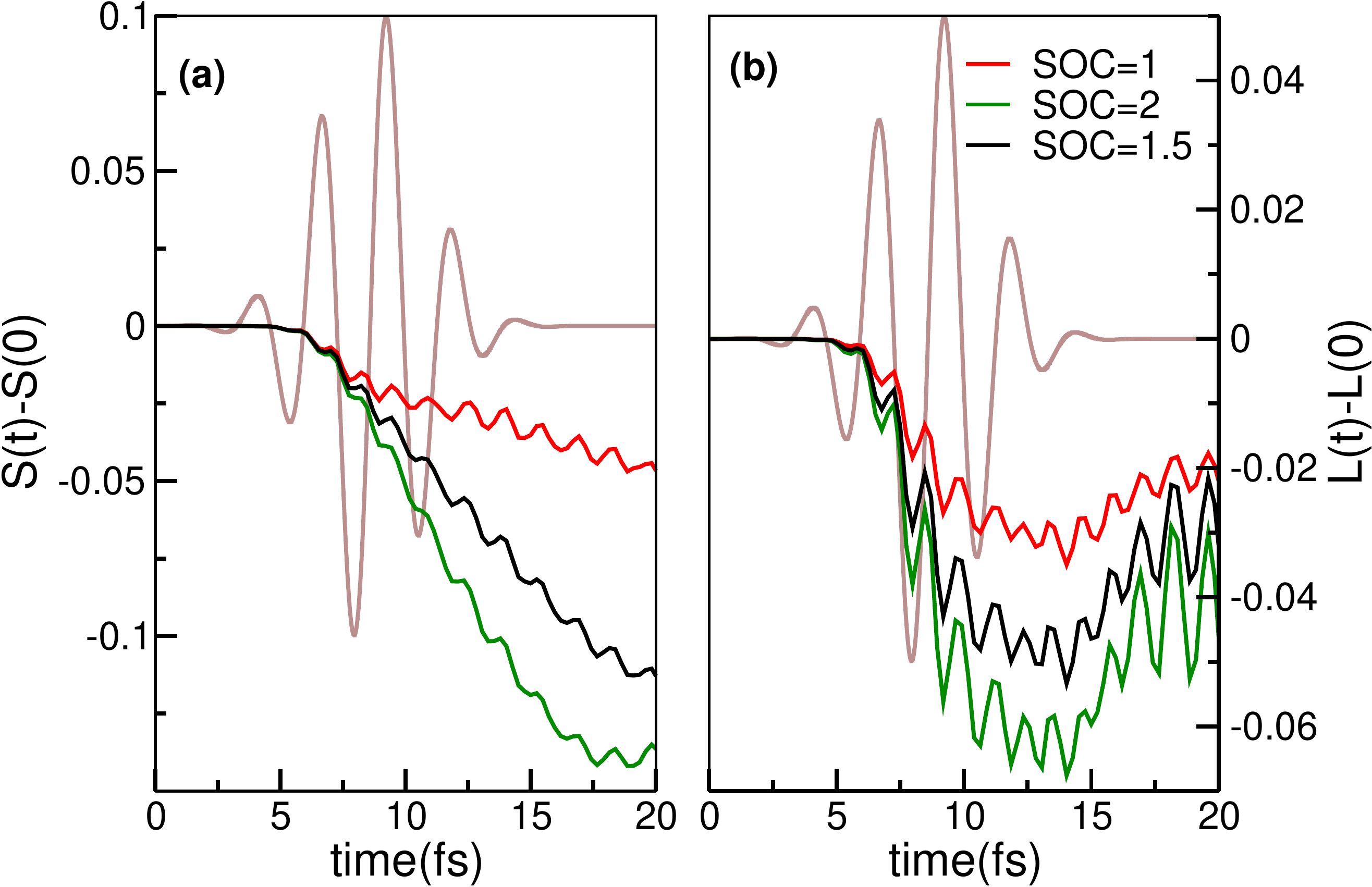}
\caption{Dependence of the dynamics of ${L_z}$ and ${S_z}$ angular momentum on the spin-orbit coupling (SOC) constant. Shown are the (a) the change in spin angular moment, $S_z(t)-S_z(t=0)$, and (b) the change in orbital angular momentum, $L_z(t)-L_z(t=0)$ for Co, with the $A$-field of the pump pulse is also presented. Increasing the SOC results both in an increased loss of ${ L_z}$ during the pulse, but also a subsequently larger increase after the pulse as the enhanced SOC drives increased ${S_z}$ transfer into ${L_z}$.} \label{fig:soc}
\end{figure}
{\it Short pulse}: change in angular momentum is driven by optical excitations and thus governed predominantly by an extrinsic time scale, that of the full width half maximum of the laser envelope.
In contrast, the loss of spin angular momentum -- driven by spin-orbit induced spin flips -- is to a large extent governed by an intrinsic time scale, the spin-orbit coupling time. Note that the spin-orbit coupling itself is time-dependent and hence depends to some extent upon the pump pulse, but is a highly material dependent property and hence has an intrinsic time scale associated with it.

By reducing the pulse full width half maximum we may therefore separate the time scales of dynamics of ${\v L}$ and ${\v S}$, and so disentangle the essential mechanism of the magnetic moment loss into the lattice.
In Fig.~\ref{fig:shortp} we show the spin dynamics of moment loss in Ni and Co for a pulse of FWHM 5~fs. While the excited charge stabilises after approximately 11~fs, the expectation values of ${L_z}$ and ${S_z}$ continue to change, with a considerable loss in spin angular momentum occurring between 15 and 30~fs. This, however, is accompanied by a pronounced increase in the orbital angular momentum of the electron system, especially clear for the Co system. The particular clear behaviour for Co can be attributed to the much longer SO time as compared to Ni, resulting in an especially clear separation of time scales. That the SO time for Co is longer than for Ni has been demonstrated before, both experimentally as well as theoretically\cite{SSBM17}.
By separating time scales we can clearly see an increase in ${L_z}$ as the material demagnetises and, as no other angular momentum channels are active at these ultrashort time scales, can conclude that this occurs solely due to transfer of ${S_z}$ angular momentum to  ${L_z}$ angular momentum. However, since the ${L_z}$ already substantially decreases much before the onset the dynamics of $S_z$, even for these pulses we will not see $L_z$ increase over its ground-state value.

The SO time can be controlled by artificial enhancement or suppression of the ground state spin orbit coupling (SOC) constant. This will control the rate of decrease of ${\v S}$, as spin-flip processes will be enhanced, however it will also change the initial state by increasing ground-state ${\v L}$. In the picture of the femtosecond dynamics of angular momentum presented here, an increase in the loss rate of ${\v S}$ due to artificially enhanced SOC must be accompanied by a corresponding increase in the gain rate of ${\v L}$. Manipulation of the SOC constant thus provides a rigorous test of our picture of the dynamics.

The dynamics of ${S_z}$ (Fig.\ref{fig:soc}a) and ${L_z}$ (Fig.\ref{fig:soc}b) in Co are shown for SOC parameter scaled by factors of 1.5 and 2.0. The increase in the ground-state value of the orbital angular momentum results in an increase in the very short time scale loss of ${L_z}$ ($t < 10$~fs), with the subsequent rate of spin demagnetisation increased due to enhancement of ultrafast SO induced spin flips. However, post pulse peak we see that this increase is matched by a corresponding increase in the orbital angular momentum, exactly as expected in our picture of the short time dynamics.
It should be noted that in both Fig.~\ref{fig:shortp} and Fig.~\ref{fig:soc} the loss of ${S_z}$ does not correspond precisely to the increase in ${L_z}$; only the total angular momentum (orbital+spin+lattice) is conserved and ${L_z}$ is continuously transferred to the lattice during the dynamics.


{\it Conclusions}: for experimentally relevant low fluence pulses (incident fluence of 2-5~mJ/cm$^2$) we have shown that excellent agreement exists for spin and orbital angular momentum obtained either (i) from the optical spectra at L-edge via sum rules (the experimental method of choice) or, (ii), directly from the muffin-tin expectation values of the corresponding operators.
On this basis we address the femtosecond dynamics of orbital and spin angular momentum, and in particular the long standing question of the mechanism by which ${\v S}$ is transferred to the lattice at ultrashort time scales. The dynamics of  ${\v L}$ are dominated by optical excitations, while that of ${\v S}$  by spin-orbit induced spin flips, endowing these momenta with distinct time scales. By employing an ultrashort pulse we show these time scales can be separated, thus temporally disentangling the dynamics of ${\v L}$ and ${\v S}$. Our results then demonstrate conclusively that the predominant mechanism by which spin angular momentum transfers to the lattice is through the orbital angular momentum of the electronic system and at ultrashort time scales this causes  ${\v L}$ to increase. Recent efforts to generate ultrashort soft X-ray pulses both at free electron laser facilities \cite{Behrens2014,Duris2020} and high harmonic radiation sources \cite{Teichmann2016} will enable such XMCD experiments to separately address the dynamics of spin and orbital angular momentum at ultrafast time scales as suggested in the present work.

\section{Acknowledgements}
Sharma and CvKS would like to thank DFG for funding through TRR227 (project A04 and A02). Shallcross would like to thank DFG for funding through SH498/4-1 while PE thanks DFG for finding through DFG project 2059421. The authors acknowledge the North-German Supercomputing Alliance (HLRN) for providing HPC resources that have contributed to the research results reported in this paper.

%

\end{document}